\documentclass[12pt]{JHEP3}
\usepackage{amsmath}
\numberwithin{equation}{section}

\newcommand{\tmtextbf}[1]{{\bfseries{#1}}}
\newcommand{\tmtextit}[1]{{\itshape{#1}}}
\title{
Can the Hagedorn Phase Transition be explained from Matrix Model for Strings?}
\author{\\
B. Sathiapalan and Nilanjan Sircar\\
Institute of Mathematical Sciences\\
CIT Campus, Tharamani\\
Chennai 600041, INDIA\\
Email: bala, nilanjan@imsc.res.in
\\
}
\preprint{IMSC/2008/05/04 \\arXiv:0805.0076}
\maketitle
\abstract{The partition function of BFSS matrix model is studied
for two different classical backgrounds up to $1$-loop level. \ One of the backgrounds
correspond to a membrane wrapped around a compact direction
 and another to a localised cluster of $D0$-branes. \ It is shown that, there exist
phase transitions between these two configurations -
 but only in presence of an IR cut-off. \ The
low temperature phase corresponds to a string (wrapped membrane) phase
and so we call this the Hagedorn phase transition.\  While the
presence of an IR cut-off
 seemingly is only required for
perturbative analysis to be valid, the physical  necessity of such a
cut-off can be seen in the dual super-gravity side. It has been argued from
entropy considerations that a finite size horizon must develop even in an extremal
 configuration of D0-branes, from
higher derivative $O(g_s)$ corrections to super-gravity.   It can then be shown
that the Hagedorn like transition exists in super-gravity also.
\ Interestingly
the perturbative analysis also shows a second phase transition back
to a string phase. This  is reminiscent of the Gregory-Laflamme instability.}

\begin{document}
\section{Introduction}

\tmtextit{Hagedorn Temperature}~{\cite{Hagedorn}} is the ``limiting
temperature'' at which the string partition function diverges due to an
exponential growth in the density of states, which overtakes the Boltzmann
suppression factor. There is some evidence to interpret this temperature as a
phase transition temperature \cite{Thorn, Atick, BSath, Kogan}. 
\ Yang-Mills theories are known to have
confinement-deconfinement phase transition, and also they are known to be dual
to String Theories with $1 / N$ (for gauge group $SU (N)$) interpreted as
the string coupling constant, $g_s$. It is then natural to identify this
Hagedorn Transition temperature with confinement-deconfinement transition
\cite{Barbon,Kalyan,Minwalla3,Sundborg,Orselli1,Orselli2,Orselli3,Orselli4}. \
Above the deconfinement transition the gluon flux tube disintegrates. \
Correspondingly one would expect that the string disintegrates above the
Hagedorn phase transition and is replaced by something else - perhaps a black
hole.

It is difficult to study the disintegration of string theory using the
perturbative string formalism. One needs a non-perturbative description where
a string can be described in terms of some other entity. One such description
is the BFSS matrix model. In this model one can construct a classical
configuration that looks like a membrane. As a 10-dimensional object it is a
D2 brane of IIA string theory. If one of the space dimensions is compactified,
the D2 brane wrapped around it,  is T-dual to a D-string (D1 brane) of IIB
string theory. This in turn is S-dual to an F(fundamental)-string. \ So on the
one hand we can pretend that this D-string is the fundamental string \ whose
phase transition we are interested in, and on the other hand this D-string is (T-dual to ) a
composite of D0 branes (arranged in a very specific way). At the phase
transition this classical membrane configuration can be expected to
disintegrate so that we end up with just a bunch of localised D0-branes. This
is thus the ``S-dual'' of the Hagedorn transition. This was what was
investigated in in a qualitative way in~\cite{Bala} . \ 
It was shown by computing (in a high temperature approximation) the one
loop free energy, that there is a phase transition from a membrane phase to a
clustered phase. However an IR cutoff was crucial for the calculation. The
motivation for the IR cutoff is roughly that the $D0$ branes are actually bound
(albeit marginally) and so one expects them to be localised. \ 
Two $D0$ brane potential, at finite temperature, studied in \cite{Bal,Ambjorn} shows 
possibility of bound state. 
\ Our aim in this paper is to redo the one-loop analysis carefully without approximations. \ 
The net result reaffirms the result of~\cite{Bala},  but with some modifications in the 
analytical expressions. \ Interestingly an additional phase transition is found 
{\it back to the string phase}. \ The one-loop partition function of strings from matrix model
\cite{Verlinde} was also studied extensively in \cite{Grignani1,Grignani2,Grignani3}.

Holography~\cite{Maldacena} has given some new insights into the dynamics 
and phase structure of super-symmetric Yang-Mills \cite{Minwalla1,Minwalla2,Minwalla3}. 
 \ This should also be taken into account. \   
Deformations of these theories have also been studied 
\cite{Polchinski1,Polchinski2,Porrati1,Porrati2, Warner}. \ 
In fact, motivated by the analysis in \cite{Polchinski1,Polchinski2}, the thermal
Hawking-Page phase transition\cite{Witten, Hawking} in AdS with
a ``hard wall'' \cite{Herzog} has been studied - the hard wall removes a portion of the AdS 
near $r=0$, which in the gauge theory corresponds to the IR region. In these models the
cutoff is a way to simulate a boundary gauge theory that is not conformally invariant, i.e.
confining. The cutoff radius is related to the mass parameter of the
gauge theory. It has in fact been shown \cite{Cai} that an
IR cutoff is crucial for the existence of two phases: BPS Dp-branes
(more precisely their near horizon limit which is AdS) and black
Dp-branes (AdS black hole), separated by a
finite temperature phase transition. Also \cite{Nemani} has argued (from entropy considerations)
that $O(g_s)$ higher derivative corrections to super-gravity must induce
a finite horizon to develop for a configuration of extremal D0-branes.
This also acts as an IR cutoff. \ Based on all this, the main conclusion of this paper is 
that  \ \tmtextit{an infrared cutoff needs to be included in the BFSS matrix model if it is 
to describe string theory.}

This is quite understandable from another point of view. \ Simple
parameter counting shows that the BFSS model, as it stands, cannot be equivalent
to string theory. It has only one dimensionless parameter $N$. \  The other
parameter is $g_{YM}$ - which is dimensionful and just sets the overall mass
scale. String theory describing $N$ D0 branes has two dimensionless parameter
$N$ and $g_s$, and a dimensionful parameter $l_s$. The two theories thus cannot be
equivalent, except in some limit $g_s \to 0$ (or $g_s \to \infty$ -
M-theory). 
\ For finite $g_s$ one
needs an additional parameter on the Yang-Mills side. \  Thus an IR cutoff $L_0$
introduces a second scale into the Yang-Mills theory, and thus the ratio of
the two scales is a dimensionless parameter and now the parameter counting
agrees. Furthermore, low energies in the matrix model that describes D0 branes,
corresponds in the super-gravity description to short distances, i.e. regions
near the D0 branes, where the dilaton profile corresponds to a large value of
$g_s$. If we include the low energy (IR) region in the configuration space of
the Yang-Mills (matrix model) theory, we are forced to include the effects of finite
$g_s$. \ The IR cutoff is an additional parameter that signifies our ignorance of
(large) strong coupling effects. \ With an IR cutoff we can maintain $g_s$ at a finite
(small) but non-zero value. \  What actually happens in this region (i.e. close to
D0 branes) due to strong coupling effects of large $g_s$ is not fully known,
but as mentioned above ~\cite{Nemani} has a plausible proposal based
on entropy arguments.  \ The suggestion is
that a finite size horizon develops. \ This if true, vindicates the introduction of
an IR cutoff. Note that another way of making the functional integral finite
 is to give a mass to the scalar fields - thereby removing the zero
 mode.  The matrix model corresponding to the BMN pp wave limit \cite{Berenstein} in fact has
 such a mass term with the mass being a free parameter. This is
 reminiscent of the N=1* theories studied in \cite{Polchinski2} 
and these techniques have been applied to the BMN matrix model \cite{Lin}. We
 can therefore take an agnostic attitude regarding the origin of the
 cutoff and treat it as the extra parameter necessary to match with string
 theory for finite $g_s$. It should correspond to the fact that the D0-brane bound
 state must have a finite size. This is not easy to see in the BFSS matrix model,
 in perturbation theory, because of the flat directions in the potential.  

As mentioned above there seems to be another high temperature string phase. This is reminiscent
of the $T\rightarrow 1/T$ symmetry that has been discussed by many authors 
\cite{O'Brien,Alvarez,Dienes,Chaudhuri}. 
The perturbative analysis  shows a second phase transition to a  string
phase at very high temperatures. This is reminiscent of  Gregory-Laflamme instability.
We comment on this briefly at the end. We show that the entropic arguments that motivate the 
Gregory-Laflamme transition can also be made for extremal black holes with finite size horizon
such as the Reissner Nordstrom black holes. 

This paper is organised as follows. \ Section 2 is a brief review of some
relevant aspects of the BFSS matrix model. \ Section 3 contains the perturbative
one-loop analysis. Section 4 contains the analysis using the super-gravity
dual. Section 5 contains some conclusions.

\section{M Theory}

M theory is the strong coupling limit of type $IIA$ string theory.\  In this
limit, it behaves as an eleven dimensional theory in an infinite flat space
background.\  At low energy, it behaves as a eleven dimensional super-gravity. \ It
also has membrane degrees of freedom with membrane tension
$\frac{1}{(l_p^{11})^3}$, where $l_p^{11}$ is eleven dimensional Planck
length. As mentioned in the introduction these membranes can be wrapped along
compact directions to form strings and a study of the partition function of M
theory may throw light on Hagedorn Transition, where strings are a particular
configuration or phase of more fundamental degrees of freedom.

\subsection{BFSS Matrix Model} 

M theory in infinite momentum frame (IMF) are
described by $D 0$-branes only, as proposed in \cite{BFSS}. \ The action is
given by dimensional reduction of $10$ dimensional $U (N)$ Super
Yang-Mills' (SYM) theory to zero space dimension (in $N \to \infty$ limit).
Where the space co-ordinates of $N$ $D 0$-branes are given by eigenvalues of
$N \times N$ matrices $A_i$, $i = 1, \cdots, 9$ of the dimensionally reduced
SYM.

Generally the infinite momentum frame is chosen by considering the eleventh
direction, $X^{11}$ to be compact (radius, $R_{11}$) and then subsequently
boosting the system in this direction. \ The negative and zero Kaluza-Klein modes
decouple, and can be integrated out. \ These positive Kaluza-Klein modes have
non-zero $RR$ charge from ten dimensional view point (which is type $IIA$
string theory by definition, with $g_s l_s = R_{11}$), and they are identified
as the $D 0$-branes of the theory . \ In the end, we must let $R_{11}$ and $N /
R_{11}$ tend to infinity to get uncompactified infinite momentum limit.

\subsection{DLCQ M-Theory}

In the method of Discrete Light Cone Quantisation(DLCQ)~{\cite{DLCQ}} we
compactify a light-like coordinate $X^-$ instead of $X^{11}$. This theory is
valid for any value of $N$. The idea is that in the large $N$ limit this
becomes equivalent to M-theory. For finite $N$ this is a very simple model
as shown in~\cite{Seiberg}.

\subsection{Relation between DLCQ M-theory and the BFSS matrix model}
We will review Seiberg's~\cite{Seiberg} arguments on the relation between
DLCQ M-Theory and BFSS Matrix Model.

In DLCQ, we compactify a light-like circle which corresponds to,
\begin{equation} \label{light_like}
\left( \begin{array}{c}
X^{11} \\
X^0 
\end{array} \right) \sim 
\left( \begin{array}{c}
X^{11} \\
X^0 
\end{array} \right) +
\left( \begin{array}{c}
\frac{R^-}{\sqrt{2}} \\
-\frac{R^-}{\sqrt{2}}
\end{array} \right)
\end{equation}
where $X^{11}$ is the longitudinal space-like direction and $X^0$ is the time-like direction 
in the $11$ dimensional space-time. \ We can consider it as a limit of compactification
on a space-like circle which is almost light like
\begin{equation} \label{almost_light_like}
\left( \begin{array}{c}
X^{11} \\
X^0 
\end{array} \right) \sim 
\left( \begin{array}{c}
X^{11} \\
X^0 
\end{array} \right) +
\left( \begin{array}{c}
\sqrt{\frac{R^-}{2} + R_{11}^2}\\
-\frac{R^-}{\sqrt{2}}
\end{array} \right)
\simeq
\left( \begin{array}{c}
X^{11} \\
X^0 
\end{array} \right) +
\left( \begin{array}{c}
\frac{R^-}{\sqrt{2}} + \frac{R_{11}^2}{\sqrt{2} R^-}\\
-\frac{R^-}{\sqrt{2}}
\end{array} \right)
\end{equation}
with $R_{11} << R^-$. \ The light-like compactification (~\ref{light_like}) is
obtained from (~\ref{almost_light_like}) as $R_{11} \to 0$. \ This compactification
is related by a large boost with
\begin{equation} \label{boost}
\beta_{v} = \frac{R^-}{\sqrt{(R^{-})^2 + 2 R_{11}^2}} \simeq 1 - \Big{(}\frac{R_{11}}{R^-}
\Big{)}^2
\end{equation}
to a spatial compactification on
\begin{equation} \label{space_like}
\left( \begin{array}{c}
X^{11'} \\
X^{0'} 
\end{array} \right) \sim 
\left( \begin{array}{c}
X^{11'} \\
X^{0'} 
\end{array} \right) +
\left( \begin{array}{c}
R_{11} \\
0
\end{array} \right)
\end{equation}
where prime denotes boosted coordinates.

 The  longitudinal boost of the light-like
circle (eqn.(~\ref{light_like})) rescales  the value of the radius of compactification, 
$R^{-'}=\delta^{-1} R^-$, where $\delta = \sqrt{\frac{1+\beta_v}{1-\beta_v}}$. \ It also
rescales the value of light-cone energy $P^-$ similarly. \ Therefore $P^-$ is proportional
to $R^-$ i.e. $P^- \sim R^- M_p^2$. The factor of $M_p =
(l_p^{11})^{- 1}$, the 11 dimensional Planck mass is introduced on dimensional
grounds. \ For small $R_{11}$, the value of $P^-$ in the system with the almost light-like circle
 is also proportional to $R^-$ (an exception to that occurs when
$P^-=0$ for the light-like circle; then $P^-$ can be non-zero for the
almost light-like circle). \  The boost (eqn.(\ref{boost})) rescales $P^-$ to be
independent of $R^-$ and of order $R_{11}$ (if originally $P^-=0$, the
resulting $P^-$ after the boost can be smaller than order $R_{11}$).\ So,
$P^{' -} \sim R_{11} M_p^2$, where $R^{-'}= R^{11} = R^- / \delta$.

Let us consider M theory compactified on light-like circle (eqn.(~\ref{light_like})) as the
$R_{11} \to 0$ limit of the compactification on an almost light-like circle 
(eqn.(~\ref{almost_light_like})) or as the limit of boosted circle (eqn.(~\ref{space_like})).
\ Notice  $R_{11} \to 0$ corresponds to a large boost, $\beta_v \to 1$ and $\delta \to \infty$.
\ The analysis shows that, the DLCQ of M theory~\cite{DLCQ} is related to the compactification 
on a small spatial circle i.e. the BFSS Matrix model~\cite{BFSS}. \ For small $R_{11}$ the theory
compactified on (~\ref{space_like}) is weakly coupled string theory with string coupling
$g_s=(R_{11} M_p)^{\frac{3}{2}}$ and string length $l_s^2 =( R_{11}M_p^3)^{-1}$. \ We see in the 
limit $R_{11} \to 0$ the string length $l_s \to \infty$, which yields a complicated theory. \ 
However $P^{' -}$ also goes to zero (if $P^-$ is initially of $O (1)$), so we are 
only interested in very low ``energy'' states of the boosted theory, 
and this simplifies things. \  This
can be made clear by rescaling parameters. \ We
have to replace DLCQ M theory by another M theory, referred to as $\widetilde{M}$
theory (BFSS Matrix theory) with Planck mass $\widetilde{M_p}$ compactified on the
spatial circle of radius $R^{11}$. \ The relations between parameters of the two
theories is obtained by keeping $P^{\prime -} \sim R^{11} \widetilde{M_p}^2$
fixed with the limit $R^{11} \to 0$ and $\widetilde{M_p} \to \infty$, we get,
\begin{equation}
  R^{11} \widetilde{M_p}^2 = R^- M_p^2
\end{equation}
And as boost does not affect transverse directions,
\begin{equation}
  M_p R_i = \widetilde{M_p} \widetilde{R_i}
\end{equation}
where $R_i$ are any length parameter in transverse direction. So,
\begin{eqnarray}
  \frac{\widetilde{M_p}}{M_p} & = & \delta^{1 / 2} \nonumber\\
  \frac{\widetilde{g_s}}{g_s} & = & \delta^{- 3 / 4} \nonumber\\
  \frac{\tilde{\alpha}'}{\alpha'} & = & \delta^{- 1 / 2} 
\end{eqnarray}
So with finite $R^-$ and $M_p$, the corresponding string theory in BFSS model
is weakly coupled and with very large string tension (Notice as $R_{11} \to 0$ or 
 $\delta \to \infty$, both $\widetilde{g}_s$ and $\widetilde{l}_s$ goes to zero). 
\ So this theory is simple.

If we compactify one of the transverse direction with radius $R_i$ and consider the
T dual along this direction, the T dual radius is given by $R_i^*=\frac{\alpha'}{R_i}$. \
So the scaling gives,
\begin{equation}
\widetilde{R}_i^*=R_i^*.
\end{equation}

In this paper we will calculate partition function of the DLCQ theory using
the BFSS matrix  model. Thus we use  parameters (denoted with tilde) which are
related to that of DLCQ by a scaling, as discussed above.

\subsection{The BFSS Matrix Model Action} 
The bosonic part of the action for $N$ $D0$-branes is given by,
\begin{equation}
  S = \frac{1}{2 \widetilde{g}_s} \int \frac{dt}{\widetilde{l}_s} Tr 
 \lbrace (D_t X^i)^2 + \frac{1}{4 \pi^2 \widetilde{l}_s^4} \lbrack X^i, X^{j} \rbrack^2 \rbrace
\end{equation}
where $D_t = \partial_t + iA_0$. \ Which is basically 10d U(N) SYM action 
(with $A^i = \frac{X^i}{2 \pi \widetilde{l}_s^2}$) 
reduced to 1d, with 1d $\widetilde{g}_{YM}^2 = \frac{1}{4 \pi^2} \frac{\widetilde{g}_s}
{\widetilde{l}_s^3}$.

The parameters of DLCQ theory  are radius $R^-$ and the eleven
dimensional Planck length $l_p^{11}$. \ While the parameters of Matrix Model are
$R^{11}$ and $\widetilde{l}_p^{11}$. \ 
The membrane tension is given by
$\frac{1}{(2 \pi)^2 (\widetilde{l}_p^{11})^3}$. \ This fixes $\widetilde{g}_s^2 = (
\frac{R^{11}}{\widetilde{l}_p^{11}})^3$ and also 
$\widetilde{\alpha}' = \frac{(\widetilde{l}_p^{11})^3}{R^{11}}$. \ These
relations also imply that $\widetilde{g}_s \widetilde{l}_s = R^{11}$. \ 
Where $\widetilde{g}_s$ is the Type $IIA$ string
coupling constant and $2 \pi \widetilde{\alpha}'$, the inverse string tension with
$\widetilde{\alpha}' = \widetilde{l}_s^2$. \ These parameters are related to that
of DLCQ by the scaling discussed. 

As mentioned in the introduction, this parameter count in the matrix
model is misleading. Written in terms of $A_\mu$ it clearly has only
one dimensionless parameter $N$ and one scale, set by $g_{YM}$. So the
dynamics depends only on $N$, i.e. all physical quantities will scale
with the appropriate power of $g_{YM}$ times some (dimensionless) function of $N$.

\subsection{Construction of membranes and strings}

In matrix theory a membrane is described in the large $N$ limit by the
configuration \cite{BFSS, Banks}
\begin{equation}
  X^i = \widetilde{L}^i p  \ , \qquad X^j = \widetilde{L}^j q
\end{equation}
where $p$ and $q$ are matrices satisfying,
\begin{equation}
 \lbrack p, q \rbrack  = \frac{2 \pi i}{N}
\end{equation}
We will consider the configuration given by
\begin{equation}
  X^9 = \widetilde{L}^9 p
\end{equation}
which describes a membrane wrapped around $X^9$ with other edges free. \ If we
consider $X^i, i \ne 9$ as periodic functions of $q$, all of form $\exp
(imq)$, then we get a closed string. \ Let us construct this string action in
matrix model.

Let us assume that $X^9$ is a compact dimension of radius $\widetilde{L}^9$, assumed to be
small. \ When a membrane is wrapped around $X^9$ we get a D string(to get F
string we have to wrap it around the $X^{11}$) with inverse tension $\widetilde{\beta}' =
\frac{(\widetilde{l}_p^{11})^3}{\widetilde{L}^9}$, and string coupling $\widetilde{g}_{s \beta'} =
\frac{(\widetilde{L}^9)^3}{(\widetilde{l}_p^{11})^3}$.

As shown in~\cite{Bala} , matrix model action, at zero temperature, with background
configuration given by membranes constructed in the above way, matches exactly
with the string theory action in light cone frame. \ The bosonic part of the
action used in this paper is given by
\begin{equation}\label{action_bala}
  S = \frac{1}{2 \widetilde{g}_s} \int \frac{dt}{\widetilde{l}_s} 
  \int_0^{2 \pi \widetilde{L}_9^*}
  \frac{dx}{2 \pi \widetilde{L}_9^*} Tr\{(D_t X^i)^2 - (D_x X^i)^2 + (F_{09})^2 +
  \frac{1}{4 \pi^2 \widetilde{l}_s^4} \lbrack X^i, X^j \rbrack^2 \}
\end{equation}
Which is a $1 + 1$ dimensional action, obtained by dimensional reduction of $9
+ 1$ dimensional $U (N)$ Super-symmetric Yang-Mills action and subsequently
taking T-dual along the compact direction $X_9$ of radius $\widetilde{L}^9$. $D_x =
\partial_x + iA_9$ is the covariant derivative in a direction $X_9^*$
,which is T-dual to $X_9$, and has a radius $\widetilde{L}_9^{\ast} =
\frac{\widetilde{\alpha}'}{\widetilde{L}_9}$. 
$x$ is the co-ordinate along a $D$$1$ brane wound around $X_9^{\ast}$.

Following Taylor's~{\cite{taylor}} calculation we have,
\begin{equation}
  A^9 = \frac{1}{2 \pi \widetilde{\alpha}'} \sum_{n = - \infty}^{\infty} \exp (inx
  \frac{L_9}{\widetilde{\alpha}'}) X_{0 n}^9
\end{equation}
Where $X_{00}^9 = \widetilde{L}^9 p$ is the original $D$$0$ brane matrix of uncompactified
theory. \ Thus $D_x$ is given by
\begin{equation}
  D_x = \partial_x \otimes I + I \otimes \frac{\widetilde{L}^9}{\widetilde{\alpha}' N} \partial_q
\end{equation}
which acts on eigen-functions
\begin{equation}
\label{eigen}
  e^{ir \frac{x}{\widetilde{L}_9^{\ast}}} e^{imp} e^{inq}
\end{equation}
with eigen values $\frac{rN + n}{N\widetilde{L}_9^{\ast}}$. \ This action(\ref{action_bala}), 
in the large $N$ limit, matches with closed string action~\cite{Bala}, where the effective radius
of world-sheet is $N\widetilde{L}_9^{\ast}$ and inverse string tension is 
$2 \pi \widetilde{\beta}'$, $\widetilde{\beta}' = \widetilde{L}_9^{\ast} R^{11}$. \ 
Turning on $F_{09}$ corresponds to addition of F-strings. \ The commutator terms
are zero if we restrict the matrices $X^i$ to be $X^i (x, q, t)$ i.e. without
any $p$ dependence. \ $p$ dependence corresponds to fluctuations in matrix model
that are not string-like.

\subsubsection{Two Phases}

\tmtextbf{Phase 1:} The background $X^9 = \widetilde{L}^9 p$ gives a configuration where
the $D0$ branes spread out to form a string wound in the compact direction.\\
\tmtextbf{Phase 2:}The background $X^9 = 0$ gives a phase where the $D0$-branes are clustered.

We will consider these two backgrounds to calculate free energy up to one loop
level, and compare to find any signature of phase transition. \ It is important to
have a precise definition of the measure in the functional integral. \ This is 
described in Appendix A.

\section{One Loop Free Energy}
For convenience we will drop the tilde sign on the parameters and put it back in the end.

The details are given in the Appendices B and C. We summarise the
results here. \ We will first calculate action for a SUSY scalar field on $S^1 \times S^1$,
 which is then related to the action (\ref{action_bala}) we are concerned in the 
following subsection. \ We start with the bosonic part: \\
Consider the Euclidean action,
\begin{equation}
  S = \frac{1}{g_s} \int_0^{\beta} \frac{dt}{l_s} \int_0^{2 \pi L_9^{\ast}}
  \frac{dx}{2 \pi L_9^{\ast}} \{ (\partial_t X)^2 + (\partial_x X)^2 \}
\end{equation}
Where $t$ and $x$ directions are both periodic with periods $\beta$ and $2 \pi
L_9^{\ast}$ respectively, then,
\begin{equation}
  X = \sqrt{2 \pi L_9^{\ast} \beta} \sum_{n = - \infty}^{\infty} \sum_{m = -
  \infty}^{\infty} X_{nm} e^{- \frac{2 \pi i}{\beta} nt} e^{-
  \frac{i}{L_9^{\ast}} mx}
\end{equation}
$X$ is real implies $X_{nm}^{\ast} = X_{- n - m}$.So we get,
\begin{equation}
  S = \frac{\beta}{2 g_s l_s} \sqrt{2 \pi L_9^{\ast} \beta} \sum_{n = -
  \infty}^{\infty} \sum_{m = - \infty}^{\infty} \lbrack ( \frac{2 \pi n}{\beta})^2 + (
  \frac{m}{L_9^{\ast}})^2 \rbrack X_{nm} X_{- n - m}
\end{equation}
Now we can calculate partition function easily (see Appendix B). \ We get,
\begin{equation}
  Z = \frac{L_0}{\sqrt{2 \pi g_s l_s \beta}} \eta (i \frac{\beta}{2 \pi
  L_9^{\ast}})^{- 2}
\end{equation}
where $\eta (x)$ is Dedekind's Eta function and $L_0$ is cut-off introduced in
zero mode integral.\\
The Partition function in terms of ratio of radii of two $S^1$, ie. $x =
\frac{\beta}{2 \pi L_9^{\ast}}$,
\begin{equation}
  Z = \frac{L_0}{\sqrt{4 \pi^2 g_s l_s L_9^{\ast}}} \frac{1}{\sqrt{x}} \eta
  (ix)^{- 2}
\end{equation}
Dedekind's Eta Function has a symmetry given by,
\begin{equation}
  \eta (ix) = \frac{1}{\sqrt{x}} \eta (i / x)
\end{equation}
Which makes the partition function invariant under the transformation $x \to 1
/ x$

For low temperature, $\frac{\beta}{2 \pi L_9^*} >> 1$, the free energy takes the form,
\begin{equation}
F(T) = - \frac{1}{\beta} ln(Z) \simeq -\frac{1}{12 L_9^*} - \frac{1}{2} T \ 
ln(\frac{L_0^2}{2 \pi g_s l_s} T) 
\end{equation}
which shows $F(0) \neq 0$ due to the presence of zero-point energy,
\begin{equation}
F(0) =  -\frac{1}{12 L_9^*} = \sum_{n=1}^{\infty} \frac{n}{ L_9^*}
\end{equation}
using Zeta function regularization. \ The high temperature expansion, 
$\frac{\beta}{2 \pi L_9^*} << 1$ is given as,
\begin{equation}
F(T) = -\frac{\pi^2 L_9^* T^2}{3} + \frac{T}{2} \ ln(\frac{8 \pi^3 g_s l_s L_9^{*2}}{L_0^2} T)
\end{equation}


Now we add in the fermions:

The Minkowaski action is given by,
\begin{equation}
  S_M = \frac{- i}{g_s} \int \frac{dt_M}{l_s} \int_0^{2 \pi L_9^{\ast}}
  \frac{dx}{2 \pi L_9^{\ast}} \{ (\partial_{t_M} X)^2 - (\partial_x X)^2 +
  \bar{\psi} (i \gamma^{\mu}) \partial_{\mu} \psi \}
\end{equation}
$\psi_{\alpha}$, $\alpha = 1, 2$ are two components (real) of two dimensional
Maiorana Spinor $\psi$.\\
$\gamma$ matrices are given by,
\begin{equation}
  \gamma^0 = \left( \begin{array}{cc}
    0 & i\\
    - i & 0
  \end{array} \right), \gamma^1 = \left( \begin{array}{cc}
    0 & i\\
    i & 0
  \end{array} \right)
\end{equation}
\begin{equation}
  \gamma^{\mu}, \gamma^{\nu} = 2 g^{\mu \nu}, g^{00} = - g^{11} = 1
\end{equation}
Fermionic part of the action can be rewritten as,
\begin{equation}
  \bar{\psi} (i \gamma^{\mu}) \partial_{\mu} \psi = i \psi_1
  (\partial_{t_M} - \partial_x) \psi_1 + i \psi_2 (\partial_{t_M} +
  \partial_x) \psi_2
\end{equation}
Now to go to Euclidean action at finite temperature we have to take $t_M \to
it$,and $t$ compact with periodicity $\beta$.We get,
\begin{equation}
  S = - \frac{1}{g_s} \int_0^{\beta} \frac{dt}{l_s} \int_0^{2 \pi L_9^{\ast}}
  \frac{dx}{2 \pi L_9^{\ast}} \{ (\partial_t X)^2 + (\partial_x X)^2 - \psi_1
  (\partial_t - i \partial_x) \psi_1 - \psi_2 (\partial_t + i \partial_x)
  \psi_2 \}
\end{equation}
Where $X$ is periodic in both $t$ and $x$, $\psi_{\alpha}$ is anti-periodic in
$t$ and periodic in $x$.
\begin{equation}
  X = (2 \pi L_9^{\ast} \beta)^{1 / 2} \sum_{n = - \infty}^{\infty} \sum_{m =
  - \infty}^{\infty} X_{nm} e^{- \frac{2 \pi i}{\beta} nt} e^{-
  \frac{i}{L_9^{\ast}} mx}
\end{equation}
\begin{equation}
  \psi_{\alpha} = (2 \pi L_9^{\ast} \beta)^{1 / 4} \sum_{n = - \infty, n =
  odd}^{\infty} \sum_{m = - \infty}^{\infty} \psi_{\alpha, nm} e^{- \frac{\pi
  i}{\beta} nt} e^{- \frac{i}{L_9^{\ast}} mx}
\end{equation}
$X$ and $\psi_{\alpha}$ is real implies $X_{nm}^{\ast} = X_{- n - m}$ and
$\psi_{\alpha, nm}^{\ast} = \psi_{\alpha, - n - m}$. \ $X_{nm}$ is a
dimensionless c-number and $\psi_{\alpha, nm}$ is a dimensionless Grassmann
number.\\
So the action becomes,
\begin{eqnarray}
  & S = - \frac{\beta}{2 g_s l_s} \sum_{n = - \infty}^{\infty} \sum_{m = -
  \infty}^{\infty} (2 \pi L_9^{\ast} \beta) \lbrack ( \frac{2 \pi n}{\beta})^2 + (
  \frac{m}{L_9^{\ast}})^2 \rbrack X_{nm} X_{- n - m} \nonumber\\
  & - \frac{\beta}{2 g_s l_s} \sum_{n = - \infty, n = odd}^{\infty} \sum_{m =
  - \infty}^{\infty} \{ i \sqrt{( 2 \pi L_9^{\ast} \beta)} \lbrack \frac{\pi n}{\beta} +
  i \frac{m}{L_9^{\ast}} \rbrack \psi_{1, nm} \psi_{1, - n - m} + \nonumber\\
  & i \sqrt{( 2 \pi L_9^{\ast} \beta)} \lbrack \frac{\pi n}{\beta} - i
  \frac{m}{L_9^{\ast}} \rbrack \psi_{2, nm} \psi_{2, - n - m} \}
\end{eqnarray}
For Bosonic part of the action we will get same partition function ($Z_B$) as
previous section. \ Each $\psi_{\alpha}$ contributes same amount to partition
function, $Z_F = Z_{F 1} Z_{F 2} = Z_{F 1}^2 = Z_{F 2}^2$. \ We get (see Appendix C),
\begin{equation}
  Z_F = Z_{F \alpha}^2 = 2 \frac{\eta (2 ix)}{\eta (ix)}^2
\end{equation}
Where $x = \frac{\beta}{2 \pi L_9^{\ast}}$.\\
So SUSY partition function,
\begin{equation}
  Z = Z_B Z_F = Z = \frac{2 L_0}{\sqrt{4 \pi^2 g_s l_s L_9^{\ast}}}
  \frac{1}{\sqrt{x}} \frac{\eta^2 (2 ix)}{\eta^4 (ix)}
\end{equation}
Compare with,
\begin{equation}
  Z_B = \frac{L_0}{\sqrt{4 \pi^2 g_s l_s L_9^{\ast}}} \frac{1}{\sqrt{x}} \eta
  (ix)^{- 2}
\end{equation}
Now $Z$ does not have the $x \to 1 / x$ symmetry, which is natural as two
directions are not similar due to different boundary condition. \ At low
temperature, i.e. $x \to \infty$, $Z = 2 L_0 \sqrt{\frac{( \frac{1}{g_s l_s})}{2
\pi \beta}}$, which is the partition function for a super-symmetric free
particle (the zero mode).

\subsection{Free energy for two phases of Matrix Model} 
We use Background Gauge Fixing Method (see Appendix D) to calculate the free energy up to 
one loop for the action (\ref{action_bala}). 
\ The ghost terms effectively cancels the two gauge fields, and remaining theory is
effectively that of $8$ SUSY scalar fields, except the fields are now $U (N)$
matrices in adjoint representation and the derivatives are little complicated
than that for scalar fields. \ Now in phase 2 (clustered) the covariant derivatives
reduces to ordinary derivatives. \ The partition function for phase 2 (clustered)
is,
\begin{eqnarray}
 Z_2 & = & e^{- \beta F_2} = \Big{\{} \frac{2 \widetilde{L}_0}{\sqrt{4 \pi^2 \widetilde{g}_s 
\widetilde{l}_s \widetilde{L}_9^{\ast}}}
\frac{1}{\sqrt{x}} \frac{\eta^2 (2 ix)}{\eta^4 (ix)} \Big{\}}^{8 N^2} \\
\beta F_2 & = & - 8 \ N^2 \ ln (b) + 8 \ N^2 \ ln ( \sqrt{x}) - 16 \ N^2 \ ln (\eta (2 ix))
\nonumber  \\
&+& 32 \ N^2 \ ln (\eta (ix)) 
\end{eqnarray}
The $N^2$ comes as the eigenvalues are independent of $m$ and $n$ (see eqn(~\ref{eigen})). \ 
For phase 1 (string) the background of $A_9$ effectively changes the radius 
$\widetilde{L}_9^{\ast}$ to $N \ \widetilde{L}_9^{\ast}$ (see section ($2.5$)), and the 
partition function for phase 1 is given by,
\begin{eqnarray}
Z_1 & = & e^{- \beta F_1} = \Big{\{} \frac{2 \widetilde{L}_0}{\sqrt{4 \pi^2 \widetilde{g}_s 
\widetilde{l}_s N \widetilde{L}_9^{\ast}}}
\frac{1}{\sqrt{\frac{x}{N}}} \frac{\eta^2 (i \frac{2 x}{N})}{\eta^4 (i
\frac{x}{N})} \Big{\}}^{8 N} \\
\beta F_1 & = & - 8 \ N \ ln (b) + 8 \ N \ ln ( \sqrt{x}) - 16 \ N \ ln (\eta (i \frac{2 x}{N}))
\nonumber  \\ 
&+& 32 \ N \ ln (\eta (i \frac{x}{N})) 
\end{eqnarray}
The $N$ comes as eigenvalues are independent of $m$ (see eqn(~\ref{eigen})).\ 
Where $x = \frac{\beta}{2\pi \widetilde{L}_9^{\ast}}$, $b = \frac{2 \widetilde{L}_0}
{\sqrt{4 \pi^2 \widetilde{g}_s \widetilde{l}_s \widetilde{L}_9^{\ast}}}
 = \frac{2 \widetilde{L}_0}{\sqrt{4 \pi^2 \widetilde{\beta}'}}$.

\subsubsection{Low temperature expansion} 
Consider $x > > 1$ and $x / N > >1$,
\begin{eqnarray}
  \beta F_1 & \simeq & - 8 Nln (b) + 8 Nln ( \sqrt{x}) \\
  \beta F_2 & \simeq & - 8 N^2 ln (b) + 8 N^2 ln ( \sqrt{x}) 
\end{eqnarray}
Using,
\begin{equation}
  \eta (ix) \simeq e^{- \frac{\pi x}{12}} \text{ for } x > > 1
\end{equation}
So $\beta F_1 < \beta F_2$, i.e. the string phase will be favoured at low
temperature if $b / \sqrt{x} \leq 1$. \ We can expect a phase transition
from the string phase to the clustered phase as the temperature is increased from zero,
at $x = b^2$. \ For the transition temperature to lie in the validity region of the low
temperature expansion : $b > > \sqrt{N}$. \ Let us call this temperature as $T_H$.
\begin{equation}
  T_H = \frac{\pi R^{11}}{2 \widetilde{L}_0^2}
\end{equation}
In terms of  the DLCQ parameters,
\begin{equation}
  T_H = \frac{\pi R^-}{2 L_0^2}
\end{equation}
using the scaling properties discussed in section ($2.3$). \ If we now take the limit
$L_0 \to \infty$, the transition temperature $T_H \to 0$. \ \emph{So, it is essential to have
a finite value of $L_0$ to get phase transition at finite temperature.} \ This is the temperature
 at which there is a deconfinement transition in the Yang-Mills' model, which should be same
as \emph{Hagedorn transition}.

\subsubsection{High temperature expansion} 
Consider $x < < 1$,
\begin{eqnarray}
  \beta F_1 & \simeq & 8 Nln ( \frac{2 N}{b}) - 8 Nln ( \sqrt{x}) - \frac{2
  N^2 \pi}{x} \\
  \beta F_2 & \simeq & 8 N^2 ln ( \frac{2}{b}) - 8 N^2 ln ( \sqrt{x}) -
  \frac{2 N^2 \pi}{x} 
\end{eqnarray}
Using,
\begin{equation}
  \eta (ix) \simeq e^{- ( \frac{\pi}{12 x} + ln \sqrt{x})} \text{ for } x < < 1
\end{equation}
We see that, at $x > \frac{4}{b^2}$, the clustered phase is favoured but at very
high temperature we can again have a string phase. \ This ``Gregory-Laflamme'' kind of 
transition will occur
at $x = \frac{4}{b^2} N^{- 1 / N} \simeq \frac{4}{b^2}$ (For large $N$, $N^{-
1 / N} \sim 1$). \ This is also consistent with $b > > \sqrt{N}$. \ Let us call
this temperature as $T_G$.
\begin{equation}
  T_G = \frac{\widetilde{L}_0^2}{8 \pi^3 R^{11} \widetilde{L}_9^{\ast 2}}
\end{equation}
In terms of the DLCQ parameters,
\begin{equation}
 T_G = \frac{L_0^2}{8 \pi^3 R^- L_9^{\ast 2}}
\end{equation}
using the scaling properties discussed in section $2.3$. \ In this case, note
$L_0 \to \infty$ implies $T_G \to \infty$.

Let us express this in terms of the parameters of Yang-Mills theory : The infrared cutoff on
$A = {X\over \widetilde{l}_s^2} $ is ${\widetilde{L}_0\over \widetilde{l}_s^2 } = 
{1\over \widetilde{L}_0^{*}}$. Thus we get (up to factors of $2\pi$)
\begin{equation}\label{temp_GL}
T_G = {\widetilde{l}_s^4 \over R_{11} \widetilde{L}_9^{*2} \widetilde{L}_0^{*2}} = 
{\widetilde{l}_s^3 \over \widetilde{g}_s \widetilde{L}_9^{*2} \widetilde{L}_0^{*2}} =
{1\over \widetilde{g}^2_{YM 0+1}}
{1\over  \widetilde{L}_9^{*2} \widetilde{L}_0^{*2}}= {1\over \widetilde{g}^2_{YM 1+1}}
{1\over  \widetilde{L}_9^{*} \widetilde{L}_0^{*2}}
\end{equation}

\section{Review of Super-gravity Results}
The BFSS matrix model with tilde parameters is supposed to be a non-perturbative
description of M-theory ( or $IIA$ String Theory). \ But the idea of gravity
dual~\cite{Maldacena} of Yang-Mills' theories allows us to relate the matrix model
to super-gravity (with tilde parameters). One can use this to infer
properties of the matrix model.


The Yang-Mills' coupling  associated with $D0$ brane action defined in section
(2.4) is given by $g_{YM}^2=\frac{1}{4 \pi^2} \frac{g_s}{l_s^3} =\widetilde{g}_{YM}^2 
=\frac{1}{4 \pi^2} \frac{\widetilde{g_s}}{\widetilde{l_s}^3}$ is finite in the 
limit $\widetilde{g_s} \to 0$ and $\widetilde{l_s} \to 0$, as $g_s$ and
$l_s$ are finite parameters. \ This is the decoupling limit discussed in
~\cite{Maldacena} ~\cite{Seiberg} ~\cite{Sen}. \ In this limit, if we also
take $N$ large, the theory is dual to Type $II$ super-gravity solution 
discussed in~\cite{Maldacena}. \  Two classical solutions of Type $II$ 
super-gravity are : ($1$) the decoupling limit of black $D0$ branes and ($2$) the 
BPS $D0$ branes. \ In a recent study~\cite{Cai} phase transition of these two solutions were
discussed. \ It was shown that the IR cut-off plays a crucial role in
phase transition. \ As mentioned earlier this is motivated by the work
of \cite{Herzog,Polchinski1,Polchinski2}. 
\ We will redo the analysis of this paper~\cite{Cai} here and try to explain
the physical origin of IR cutoff used in~\cite{Cai} . \ The super-gravity solutions have
tilde parameters, but for  convenience we will drop the tilde signs and put them
back at the end.

Ideally, we should construct the super-gravity solution corresponding to the
wrapped membrane. \ We reserve this for the future. \ Here we are only interested
in understanding the nature of the phase transition and the role of the IR cutoff,
so we will just use the solution for $N$ coincident D0-branes used also in \cite{Cai}. \ 
In the decoupling limit, (with $U = \frac{r}{l_s^2}=$ fixed, where $r$ is radial co-ordinate
defined in the transverse space of the brane. \ $U$ also sets the energy scale of the dual 
Yang-Mills theory.) the  solution for $N$ coincident black $D0$-branes in Einstein frame is 
given by, 
\begin{eqnarray}
ds^2_{Ein}&=&\frac{\alpha'}{2 \pi g_{YM}} \Bigg{\{} -
\frac{U^{\frac{49}{8}}}{(g^2_{YM}d_0 N)^{\frac{7}{8}}}\left(1-\frac{U^{7}_H}{U^{7}}\right)dt^2
\nonumber \\ 
&+& \frac{(g^2_{YM}d_0 N)^{\frac{1}{8}}}{U^{\frac{7}{8}}}\left[\frac
{dU^2}{1-\frac{U^{7}_H}{U^{7}}}+U^2 d\Omega^2_8 \right]\Bigg{\}},
\\
e^{\phi} &=& 4 \pi g_{YM}^2 \left( \frac{g_{YM}^2 d_0 N}{U^7 } \right)^{\frac{3}{4}}, \\
F_{U0} &=&- \alpha'^{\frac{1}{2}} \frac{7 U^6 }{4 \pi^2 d_0 N g_{YM}^4}.
\end{eqnarray}
where $d_0 = 2^7 \pi^{9/2} \Gamma(7/2)$ is a constant. \ Simply setting $U_H = 0$ gives
the solution for $N$ coincident BPS D0-branes.

The Euclidean action can be obtained by setting $t=i \tau$.
The Euclidean time $\tau$ has a period
\begin{equation}
\label{bltemp}
\beta = \frac{4 \pi g_{YM} \sqrt{d_0 N}}{7 U_H^{\frac{5}{2}}}
\end{equation}
in order to remove the conical singularity. \ This is the inverse Hawking temperature of
the black $D0$ brane in the decoupling limit.

Now the on-shell Euclidean action for the two solutions can be calculated and gives,
\begin{eqnarray}
I_{black}&=& \frac{7^3}{16} \frac{V(\Omega_8) \beta}{16 \pi G'_{10}}
\int_{U_H \ or \ U_{IR}}^{U_{uv}} U^6 dU \label{On-shell-action-black}\\
I_{bps}&=& \frac{7^3}{16} \frac{V(\Omega_8) \beta'}{16 \pi G'_{10}}
\int_{U_{IR}}^{U_{uv}} U^6 dU \label{On-shell-action-bps}
\end{eqnarray}
where $G'_{10}=\alpha'^{-7} G_{10}= 2^7 \pi^{10} g_{YM}^4$ is finite in the decoupling limit. 
\ $U_{uv}$ is introduced
to regularise the action and is taken to $\infty$ in the end. \ The temperature of BPS branes
$\beta'$ is arbitrary and can be fixed by demanding the temperature of both the solutions to
be same at the UV boundary $U_{uv}$, which gives 
$\beta' = \beta \sqrt{1- \frac{U_H^7}{U_{uv}^7}} $. \ $U_{IR}$ is a IR cut-off which removes
the region $U < U_{IR}$ of the geometry. \ 
The integration in the action starts from $U_{IR}$ for BPS solution and,  
$U_{IR}$ or $U_H$ for the black brane solution depending on  $U_H < U_{IR}$ or 
$U_H > U_{IR}$ respectively.
\ If we put $U_{IR}=0$ i.e. in absence of the IR cut-off,
comparison of the actions (eqns.(~\ref{On-shell-action-black}),(~\ref{On-shell-action-bps})) 
shows that there is no phase transition, and the black brane phase is always favoured. \
Let us consider the case $U_H > U_{IR}$,
\begin{equation}
\Delta I_{bulk} = \lim_{U_{uv} \to \infty} (I_{black}-I_{bps})
= \frac{7^2}{16}\frac{V(\Omega_8) \beta}{16 \pi G'_{10}}(-\frac{1}{2} U_H^7 + U_{IR}^7) 
\end{equation}
Which shows a change in sign as we increase the temperature i.e. $U_H$ (eqn.(~\ref{bltemp})). \  
The system will undergo a phase transition (``Hawking-Page Phase transition'') 
from BPS brane to Black brane solution at $U_H^7 = 2  U_{IR}^7$. \ Actually, we should also 
consider Gibbons-Hawking surface term for careful analysis (as done in~\cite{Cai}) 
which corrects the transition temperature by some numerical factor given by,
\begin{equation}\label{Cai_crit_temp}
\beta_{crit} = \frac{4 \pi \widetilde{g}_{YM} \sqrt{d_0 N}}{7 (\frac{49}{20})^{5/14} \
\widetilde{U}_{IR}^{5/2}}
\end{equation}
We see a IR cutoff is essential to realize a phase transition,
(as $\widetilde{U}_{IR} \to 0$, $\beta_{crit} \to \infty$) so
to get confinement-deconfinement phase transition in dual super Yang-Mills theory we have to 
introduce a IR cutoff.

As mentioned in the introduction, one possible mechanism for the origin of the cutoff for 
$D0$ branes can be understood from the analysis 
of~\cite{Nemani}. \ It was shown that the higher derivative corrections to super-gravity 
introduce a finite horizon area for extremal $D0$ brane solution
which is otherwise zero. \ The multigraviton states (with total N units
of momentum in the 11th direction) and the single graviton state seem
to both be microstates of the same black hole when interaction effects
higher order in $g_s$ are included.  
\ Radius of the horizon developed due to higher derivative corrections is 
$R \sim \widetilde{l}_s \widetilde{g}_s^{1/3}$. \ So we can get 
an estimate of IR cutoff by identifying $R$ with the IR cutoff in our case, 
$\widetilde{U}_{IR}=\frac{R}{\widetilde{l}_s^2} \sim \frac{\widetilde{g}_s^{1/3}}
{ \widetilde{l}_s} \sim \widetilde{g}_{YM}^{2/3}$, which is finite in the scaling limit. \ 
If we plug in this value of $\widetilde{U}_{IR}$ in eqn.(~\ref{Cai_crit_temp}), we get
\begin{equation}
\beta_{crit} = \frac{4 \pi \sqrt{d_0 N}}{7 (\frac{49}{20})^{5/14} \
\widetilde{g}_{YM}^{2/3}} \sim \frac{1}{\widetilde{U}_{IR}}
\end{equation}

In case of $D1$ brane system, which is just T dual to the system studied above also shows
that a IR cutoff is required for phase transition~\cite{Cai}. \  We were unable to find
a analysis like~\cite{Nemani} corresponding to wrapped membrane system, which we need to 
get an estimate of the IR cutoff. 

\subsection{Gregory-Laflamme Transition}

In our calculation we find a temperature $T_G$, where the $D0$-branes spread out uniformly along
the compact space. \ This configuration is just the one that is favoured at very low 
temperatures. \ It is not clear whether this perturbative result is
reliable. \ However, a similar phase transition exists in the dual super-gravity theory,
 known as ``black hole-black string'' transition or Gregory-Laflamme transition~\cite{Minwalla1,
Minwalla2,GL1,GL2,Harmark,Hanada}. 
\ It is shown in~\cite{Minwalla1}, that the near horizon geometry of a charged black string
in $R^{8,1} \times S^1$ (winding around the $S^1$) develops a Gregory-Laflamme instability
at a temperature $T_{GL} \sim \frac{1}{L^2 g_{YM} \sqrt{N}}$, 
where $L$ is the radius of $S^1$ and $g_{YM}$ is $1+1$ dimensional Yang-Mills' coupling.
\ Below this temperature the system collapses to a black-hole. 
\ In the weak coupling limit, the dual $1+1$ SYM theory also shows 
a corresponding phase transition by clustering of eigenvalues of the gauge field 
in the space-like compact direction below the temperature, 
$T_{GL}^{'} \sim \frac{1}{L^3 g_{YM}^2 N}$, as shown numerically
in~\cite{Minwalla1}. This should be compared to the perturbative result (eqn.(~\ref{temp_GL})) \ 
$ T_{G} \sim {1\over g_{YM}^2 (L_0 ^*)^2 L_9^*}$. 
 \ So the presence of high temperature string phase in our model must correspond to some
kind of ``Gregory-Laflamme'' transition in dual super-gravity. This is
(at least superficially)
independent of the issue of any classical instability. \ This is because
both solutions may be locally stable, but at finite temperatures it is
possible to have a first order phase transition to the global minimum.

In our perturbative result, the high temperature phase is a string rather than a black string
i.e. it is the same as the low temperature phase. \ The question thus arises whether
Gregory-Laflamme transitions can happen for extremal objects. \ 
We can give a heuristic entropy argument to show Gregory-Laflamme kind of transition is
also possible for extremal system. \ In the original argument \cite{GL3}, it was shown that 
for extremal branes there is no instability. \ However, these systems had zero horizon 
area. \ Instead, we will here consider a $5$ dimensional extremal RN 
black hole with a large compact direction, and the same solution with the mass smeared
uniformly along the compact direction (``RN black ring'').
The metric, ADM mass ($M_5$) and entropy ($S_5$) for a $5d$ extremal RN black hole solution 
is given by (where the compact direction is approximated by a non-compact one),
\begin{eqnarray}
ds_5^2 &=& - (1 - \frac{r_e^2}{r^2})^2 dt^2 + (1 - \frac{r_e^2}{r^2})^{-2} dr^2 + r^2 d\Omega_3^2
\\
M_5 &=& \frac{3 \pi}{4 G_5} r_e^2
\\
S_5 &=& \frac{2 \pi^2 r_e^3}{4 G_5} = \frac{\pi^2}{2} (\frac{4}{3 \pi})^{3/2} G_5^{1/2} M_5^{3/2}
\end{eqnarray}
where $G_5$ is $5d$ Newton's constant.\ Similarly, we can write the metric for $5d$ 
extremal RN Black ring, which when dimensionally reduced gives a $4d$ extremal RN black hole. \ 
The metric, ADM mass ( $M_{(4 \times 1)}$) and entropy ( $S_{(4 \times 1)}$) is given by,
\begin{eqnarray}
ds_{(4 \times 1)}^2 &=& - (1 - \frac{R_e}{r})^2 dt^2 + (1 - \frac{R_e}{r})^{-2} dr^2 + dx^2 +
 r^2 d\Omega_2^2  \nonumber \\
&& \\
M_{(4 \times 1)} &=& \frac{R_e}{G_4} \\
S_{(4 \times 1)} &=& \frac{ 4 \pi R_e^2 \times 2 \pi L}{4 G_5} = \frac{1}{2} G_5 \frac{M^2}{L}
\end{eqnarray}
where $G_4 = \frac{G_5}{2 \pi L}$ is $4d$ Newton's constant and $L$ is the radius of
the compact direction $x$. \ If we consider $M_5 = M_{(4 \times 1)} = M$ and compare the 
entropy,
\begin{equation}
\frac{S_5}{S_{(4 \times 1)}} = \frac{16}{9} \frac{L}{r_e}
\end{equation}
So, when radius of the compact direction is greater than the radius of the $5d$ black hole
horizon, the black hole solution is entropically favoured. \ As
we increase the horizon radius, there may be phase transition when horizon size becomes of the 
order of the radius of the compact dimension, above which the ``string'' solution is 
entropically favoured. \ Our analysis is a simple entropy comparison. \ As mentioned
above, this is independent of the classical stability issue, that was studied in
detail in \cite{GL3}. \ Therefore, extremal solutions with a {\it finite horizon size} may also 
show a Gregory-Laflamme kind of transition. \ This needs further study.

\section{Conclusion}

In this paper, the finite temperature phase structure of string theory
has been studied using the BFSS matrix model which is a  $0+1$
Super-symmetric Yang-Mills (SYM) theory. \ This was first studied in
perturbation theory. \ This was actually a refinement of an earlier
calculation \cite{Bala} where some approximations were made. \ The
result of this study is that there is a finite and non zero phase
transition temperature $T_H$ below which  the
preferred configuration is where the $D0$ branes are arranged in the
form of a wrapped membrane and above which  the $D0$ branes form a
localised cluster. \ It is reasonable to identify this temperature with the
``Hagedorn'' temperature, which was originally defined for the free string. 
\ We have found that $T_H \sim {1\over L_0}$, where $L_0$ is the IR cutoff of the Yang-Mills
theory which needs to be introduced to make the calculations well defined.

The $0+1$ SYM has a dual super-gravity description. \ Here also it is seen
that in the presence of an IR cutoff there is a critical temperature
above which the BPS D0 brane is replaced by a black hole.

Simple parameter counting shows that the BFSS matrix model
needs one more dimensionful parameter if it is to be compared with
string theory,so the IR cutoff $L_0$ can be thought of as
one choice for this extra parameter. \ It makes the comparison well
defined by  effectively removing 
the strongly coupled region of the configuration space in SYM as well as
in super-gravity. \ A physical justification for this
(beyond parameter counting) comes from the work of \cite{Nemani}.
\ It is shown there that the entropy matching requires even the extremal
BPS configuration of D0-branes  to develop a horizon, due to higher
derivative string loop corrections to super-gravity. \ This is an issue
that deserves further study.

Finally, the perturbative result shows a second phase transition at a
higher temperature, back to a string like phase. \ This could be an
artifact of perturbation theory. \ On the other hand, it is very similar
to the Gregory-Laflamme instability and there is also some similarity in
the expressions obtained in \cite{Minwalla1} for the critical temperature. \ This also
requires further study.

\noindent
{\bf Acknowledgements}\\
We would like to thank Nemani V. Suryanarayana and S. Kalyana Rama for several extremely useful
discussions.

\section*{Appendix}
\appendix

\section{Defining Measure for $\mathcal{N}$ $= 2$ SUSY in $1 D$}

We define measure such that
\begin{equation}
  \int DxD \psi^{\ast} D \psi exp - \pi \int_0^{\beta} dt ( \frac{x^2}{2} +
  \psi^{\ast} \psi) = 1
\end{equation}
Where $x (t)$ is a bosonic variable,and $\psi (t)$ is its super-partner.The
SUSY transformation is given by,
\begin{equation}
  \delta x = \epsilon^{\ast} \psi + \psi^{\ast} \epsilon ; \delta \psi^{\ast}
  = - \epsilon^{\ast} x ; \delta \psi = - \epsilon x.
\end{equation}
where $\epsilon^{\ast}$ and $\epsilon$ are two infinitesimal anti commuting
parameter. From these definition we can define the measure
\begin{equation}
  DxD \psi^{\ast} D \psi \equiv dx_0 \prod_{n = 1}^{\infty} (dx_n dx_{- n}) d
  \psi_0^{\ast} d \psi_0 \prod_{m > 0} d \psi_m^{\ast} d \psi_{- m}^{\ast} d
  \psi_m d \psi_{- m}
\end{equation}
The Fermionic measure in terms of $\psi_1$ and $\psi_2$,where $\psi = \psi_1 +
i \psi_2$is given by,
\begin{equation}
  D \psi^{\ast} D \psi \equiv id \psi_{10} d \psi_{20} \prod_{m > 0} d \psi_{1
  m} id \psi_{1 m}^{\ast} d \psi_{2 m} id \psi_{2 m}^{\ast}
\end{equation}
where $x (t + \beta) = x (t)$; $x_n$ are Fourier expansion co-efficient for
$x$.$\psi_m$ are Fourier expansion co-efficient for $\psi (t)$,m runs over all
integers for periodic boundary condition,but takes only odd values for
anti-periodic boundary condition.
\begin{equation}
  x (t) = \sum_{n = - \infty}^{\infty} x_n e^{- \frac{2 \pi i}{\beta} nt}
\end{equation}
\begin{eqnarray}
  \psi (t) & = & \sum_{n = - \infty}^{\infty} \psi_n e^{- \frac{2 \pi
  i}{\beta} nt} \text{for periodic boundary condition} \\
  & = & \sum_{n = - \infty, odd}^{\infty} \psi_n e^{- \frac{\pi i}{\beta} nt}
  \text{for anti-periodic boundary condition} \nonumber \\ 
  &&
\end{eqnarray}
\subsection{Zeta Function}
We will need the following results for our calculation,
\begin{eqnarray}
\zeta(s)&=&\sum_{n=1}^{\infty} n^{-s}  \nonumber \\
\zeta(s)' &=& - \sum_{n=1}^{\infty} n^{-s} \ ln(n) \nonumber  \\
\zeta_{odd}(s) &=& \sum_{n=1,n=odd}^{\infty} n^{-s} \nonumber \\
              &=& (1- 2^{-s})\zeta(s)  \nonumber \\
\zeta_{odd}(s) &=&  - \sum_{n=1,n=odd}^{\infty} n^{-s} \ ln(n) \nonumber \\
               &=& 2^{-s} \ ln2 \ \zeta(s) + (1- 2^{-s})\zeta(s)'
\end{eqnarray}
and $\zeta(0)=-\frac{1}{2}$, $\zeta(0)'=-\frac{1}{2}ln(2\pi)$. \ Which gives
$\zeta_{odd}(0)=0$ and $\zeta_{odd}(0)'= -\frac{1}{2}ln(2)$. 
\subsection{Super-symmetric ($\mathcal{N}$ $= 2$) 1D Harmonic Oscillator }
Consider the SUSY Harmonic Oscillator at finite temperature with action,
\begin{equation}
  S = \int_0^{\beta} dt ( \frac{\dot{x}^2}{2} - \psi^{\ast} \dot{\psi} +
  \frac{x^2}{2} + \psi^{\ast} \psi)
\end{equation}
Where the SUSY transformation is given by,
\begin{equation}
  \delta x = \epsilon^{\ast} \psi + \psi^{\ast} \epsilon ; \delta \psi^{\ast}
  = - \epsilon^{\ast} ( \dot{x} + x) ; \delta \psi = - \epsilon (- \dot{x} +
  x) .
\end{equation}
With periodic boundary condition on both $x$ and $\psi$,
\begin{eqnarray}
S&=& \frac{1}{2} \beta x_0^2 + \beta \sum_{n=1}^{\infty}(1+\frac{4 \pi^2 n^2}{\beta^2})x_n x_{-n}
+ \beta \psi_0^* \psi_0 + \sum_{n=1}^{\infty}(\beta + 2 \pi i n) \psi_n^* \psi_n  \nonumber \\
&& + \sum_{n=1}^{\infty}(\beta - 2 \pi i n) \psi_{-n}^* \psi_{-n}
\end{eqnarray}
So integrating by using the measure defined the partition function is,
\begin{eqnarray}
Z &=& \int Dx \ D \psi^{\ast} D \psi \ e^{-S} \nonumber \\
&=& \sqrt{\frac{2 \pi}{\beta}} \times \prod_{n=1}^{\infty} \Big{\{} 
\frac{2 \pi}{\beta (1+\frac{4 \pi^2 n^2}{\beta^2})} \Big{\}} \times \beta \times \nonumber \\
&& \prod_{n=1}^{\infty} ( \beta + 2 \pi i n) \times \prod_{n=1}^{\infty} ( \beta - 2 \pi i n)
\nonumber \\
&=& 1
\end{eqnarray}
Using $\prod_{n=1}^{\infty} C = C^{\zeta(0)}=C^{-1/2}$, where $C$ is any constant number.

If we keep periodic boundary condition on $x$,but take anti-periodic boundary
condition on $\psi$,the Super-symmetry breaks and
\begin{eqnarray}
S&=& \frac{1}{2} \beta x_0^2 + \beta \sum_{n=1}^{\infty}(1+\frac{4 \pi^2 n^2}{\beta^2})x_n x_{-n}
+ \sum_{n=1,n=odd}^{\infty}(\beta +  \pi i n) \psi_n^* \psi_n  \nonumber \\
&& + \sum_{n=1,n=odd}^{\infty}(\beta -  \pi i n) \psi_{-n}^* \psi_{-n}
\end{eqnarray}
So integrating by using the measure defined the partition function is,
\begin{eqnarray}
Z' &=& \int Dx \ D \psi^{\ast} D \psi \ e^{-S} \nonumber \\
&=& \sqrt{\frac{2 \pi}{\beta}} \times \prod_{n=1}^{\infty} \Big{\{} 
\frac{2 \pi}{\beta (1+\frac{4 \pi^2 n^2}{\beta^2})} \Big{\}} \times  \nonumber \\
&& \prod_{m=1,odd}^{\infty} ( \beta +  \pi i m) \times \prod_{m=1,odd}^{\infty} 
( \beta -  \pi i m)
\end{eqnarray}
Now rearranging the products and using $\zeta$ function to regularise the infinite products, 
(i.e. using $\prod_{n=1}^{\infty} C=C^{-1/2}$, $\prod_{n=1}^{\infty}n = \sqrt{2 \pi}$, 
$\prod_{n=1,odd}^{\infty} C = 1$, $\prod_{n=1,odd}^{\infty}n = \sqrt{2}$) we get,
\begin{equation}
Z'=\frac{\prod_{k=0}^{\infty} ( 1+ \frac{4(\beta/2)^2}{\pi^2 (2k+1)^2})}
{\frac{\beta}{2} \prod_{n=1}^{\infty} ( 1+ \frac{(\beta/2)^2}{\pi^2 n^2})}
=\coth{\beta/2}
\end{equation}
Using,$\prod_{k=0}^{\infty} ( 1+ \frac{4x^2}{\pi^2 (2k+1)^2})=\cosh{x}$ and 
$ x \prod_{n=1}^{\infty} ( 1+ \frac{x^2}{\pi^2 n^2})= \sinh{x}$.
Also notice as $\beta \to \infty$, $Z' \to 1$ i.e. Super-symmetry is restored in the zero
temperature limit.
\section{Calculation for Massless Bosonic Field theory on $S^1 \times S^1$ }

\begin{eqnarray}
  S & = & (2 \pi L_9^{\ast} \beta) \frac{\beta}{2 g_s l_s} \sum_{n = -
  \infty}^{\infty} \sum_{m = - \infty}^{\infty} \lbrack \frac{4 \pi^2 n^2}{\beta^2} +
  \frac{m^2}{L_9^{\ast 2}} \rbrack X_{nm} X_{nm}^{\ast} \\
  & = & (2 \pi L_9^{\ast} \beta) \frac{\beta}{g_s l_s} \sum_{n = 1}^{\infty}
  \sum_{m = 1}^{\infty} \{ \lbrack \frac{4 \pi^2 n^2}{\beta^2} + \frac{m^2}{L_9^{\ast 2}} \rbrack
  \lbrack X_{nm} X_{nm}^{\ast} + X_{n - m} X_{n - m}^{\ast} \rbrack \}\nonumber\\
  &  & + (2 \pi L_9^{\ast} \beta) \frac{\beta}{g_s l_s} \sum_{n = 1}^{\infty}
  \frac{4 \pi^2 n^2}{\beta^2} X_{n 0} X_{n 0}^{\ast} + (2 \pi L_9^{\ast}
  \beta) \frac{\beta}{g_s l_s} \sum_{m = 1}^{\infty} \frac{m^2}{L_9^{\ast 2}}
  X_{0 m} X_{0 m}^{\ast} \nonumber\\
  Z & = & ( \int_{- \infty}^{\infty} dX_{00}) ( \prod_{n = 1}^{\infty}
  \prod_{m = 1}^{\infty} \int_{- \infty}^{\infty} dX_{nm} dX_{nm}^{\ast} e^{-
  (2 \pi L_9^{\ast} \beta) \frac{\beta}{g_s l_s} \lbrack \frac{4 \pi^2 n^2}{\beta^2} +
  \frac{m^2}{L_9^{\ast 2}} \rbrack  X_{nm} X_{nm}^{\ast}}) \nonumber\\
  &  & ( \prod_{n = 1}^{\infty} \prod_{m = 1}^{\infty} \int_{-
  \infty}^{\infty} dX_{n - m} dX_{n - m}^{\ast} e^{- (2 \pi L_9^{\ast} \beta)
  \frac{\beta}{g_s l_s} \lbrack \frac{4 \pi^2 n^2}{\beta^2} + \frac{m^2}{L_9^{\ast 2}} \rbrack
  X_{n - m} X_{n - m}^{\ast}}) \nonumber\\
  &  & ( \prod_{n = 1}^{\infty} \int_{- \infty}^{\infty} dX_{n 0} dX_{n
  0}^{\ast} e^{- (2 \pi L_9^{\ast} \beta) \frac{\beta}{g_s l_s} \frac{4 \pi^2
  n^2}{\beta^2} X_{n 0} X_{n 0}^{\ast}}) \nonumber\\
  &  & ( \prod_{m = 1}^{\infty} \int_{- \infty}^{\infty} dX_{0 m} dX_{0
  m}^{\ast} e^{- (2 \pi L_9^{\ast} \beta) \frac{\beta}{g_s l_s}
  \frac{m^2}{L_9^{\ast 2}} X_{0 m} X_{0 m}^{\ast}}) 
\end{eqnarray}
The zero mode integral diverges. \ So we put a cut-off $\frac{L_0}{\sqrt{2 \pi
L_9^{\ast} \beta}}$ as the value of zero mode integral.
\begin{eqnarray}
  Z & = & ( \frac{L_0}{\sqrt{2 \pi L_9^{\ast} \beta}}) \prod_{n = 1}^{\infty}
  \prod_{m = 1}^{\infty} \Big{\{} \frac{2 \pi}{(2 \pi L_9^{\ast} \beta)
  \frac{\beta}{g_s l_s} ( \frac{4 \pi^2 n^2}{\beta^2} + \frac{m^2}{L_9^{\ast
  2}})} \Big{\}}^2 \nonumber\\
  & \times & \prod_{n = 1}^{\infty} \Big{\{} \frac{2 \pi}{(2 \pi L_9^{\ast} \beta)
  \frac{\beta}{g_s l_s} \frac{4 \pi^2 n^2}{\beta^2}} \Big{\}} \times  \prod_{m = 1}^{\infty}
  \Big{\{} \frac{2 \pi}{(2 \pi L_9^{\ast} \beta) \frac{\beta}{g_s l_s}
  \frac{m^2}{L_9^{\ast 2}}} \Big{\}} 
\end{eqnarray}
Using $\prod_{n = 1}^{\infty} c = c^{\zeta (0)} = c^{- 1 / 2}$, if we take out
the constant factor $(2 \pi L_9^{\ast} \beta)$ from the products, it cancels
nicely with the factor in zero mode integral. \ These products can be rearranged to
get,
\begin{eqnarray}
 && Z = \Big{\{} L_0 \prod_{n = 1}^{\infty} \frac{2 \pi}{\frac{\beta}{g_s l_s} (
  \frac{4 \pi^2 n^2}{\beta^2})} \Big{\}}_{\text{free particle with mass} = \frac{1}{g_s
  l_s}} \times \nonumber\\
 && \Big{\{} \prod_{m = 1}^{\infty} \Big{\lbrack} \sqrt{\frac{2 \pi}{\frac{\beta}{g_s l_s} (
  \frac{m^2}{L_9^{\ast 2}})}} \prod_{n = 1}^{\infty} \frac{2
  \pi}{\frac{\beta}{g_s l_s} ( \frac{4 \pi^2 n^2}{\beta^2} +
  \frac{m^2}{L_9^{\ast 2}})} \Big{\rbrack}_{\text{SHO,mass} = \frac{1}{g_s l_s} \text{freq.}
  = \frac{m}{L_9^{\ast}}} \Big{\}}^2  \nonumber \\
&&
\end{eqnarray}
therefore,
\begin{equation}
  Z = L_0 \sqrt{(} \frac{M}{2 \pi \beta}) \prod_{m = 1}^{\infty} \Big{\{} \frac{1}{2
  \sinh (\beta \omega_m / 2)} \Big{\}}^2
\end{equation}
where,
\begin{equation}
  M = \frac{1}{g_s l_s} \text{, } \omega_m = \frac{m}{L_9^{\ast}}
\end{equation}
Using,
\begin{equation}
  \eta (ix) = \prod_{k = 1}^{\infty} (2 \sinh (\pi kx))
\end{equation}
where $\eta(z)$ is Dedekind's eta function. \  We get,
\begin{equation}
  Z = \frac{L_0}{\sqrt{( 2 \pi g_s l_s \beta)}} \ \eta ( \frac{i \beta}{2 \pi
  L_9^{\ast}})^{- 2}
\end{equation}
For low temperature, $\frac{\beta}{2 \pi L_9^*} >> 1$, the free energy takes the form,
\begin{equation}
F(T) = - \frac{1}{\beta} ln(Z) \simeq -\frac{1}{12 L_9^*} - \frac{1}{2} T \ 
ln(\frac{L_0^2}{2 \pi g_s l_s} T) 
\end{equation}
which shows $F(0) \neq 0$ due to the presence of zero-point energy,
\begin{equation}
F(0) =  -\frac{1}{12 L_9^*} = \sum_{n=1}^{\infty} \frac{n}{ L_9^*}
\end{equation}
using Zeta function regularization. \ The high temperature expansion, 
$\frac{\beta}{2 \pi L_9^8} << 1$ is given as,
\begin{equation}
F(T) = -\frac{\pi^2 L_9^* T^2}{3} + \frac{T}{2} \ ln(\frac{8 \pi^3 g_s l_s L_9^{*2}}{L_0^2} T)
\end{equation}
\section{Calculation for Fermionic part of SUSY Scalar Field theory}

\begin{eqnarray}
  S_F & = & S_{F 1} + S_{F 2} = - \frac{\beta}{2 g_s l_s} \sum_{n = - \infty,
  n = odd}^{\infty} \sum_{m = - \infty}^{\infty} \Big{\{} i \sqrt{( 2 \pi L_9^{\ast}
  \beta)} \Big{\lbrack} \frac{\pi n}{\beta} + i \frac{m}{L_9^{\ast}} \Big{\rbrack} 
  \psi_{1, nm} \psi_{1, -n - m} \nonumber\\
  &+& i \sqrt{( 2 \pi L_9^{\ast} \beta)} \Big{\lbrack} \frac{\pi n}{\beta} - i
  \frac{m}{L_9^{\ast}} \Big{\rbrack} \psi_{2, nm} \psi_{2, - n - m} \Big{\}}
\end{eqnarray}
By rearranging the sum (we have dropped the index $1$ or $2$ in $\psi$),
\begin{eqnarray}
  S_{F 1} & = & - \frac{\beta}{g_s l_s} \sqrt{2 \pi L_9^{\ast} \beta} \Big{\{} \sum_{n,
  m = 1, n = odd}^{\infty} ( \frac{\pi n}{\beta} + i \frac{m}{L_9^{\ast}})
  \psi_{nm} \ i \psi_{- n - m} \nonumber\\
  & + & \sum_{n, m = 1, n = odd}^{\infty} ( \frac{\pi n}{\beta} - i
  \frac{m}{L_9^{\ast}}) \psi_{n - m} \ i \psi - nm \nonumber\\
  & + & \sum_{n = 1, odd}^{\infty} \frac{\pi n}{\beta} \psi_{n 0} i \psi_{- n 0} \Big{\}}
\end{eqnarray}
Therefore,
\begin{equation}
  Z_{F 1} = \Big{\{} \prod_{n = 1, odd}^{\infty}  C \ \frac{\pi n}{\beta} \Big{\}}
 \Big{\{} \prod_{n = 1,odd}^{\infty} \prod_{m = 1}^{\infty} C^2 \Big{(} \frac{\pi^2 n^2}{\beta^2} +
  \frac{m^2}{L_9^{\ast 2}} \Big{)} \Big{\}}
\end{equation}
where, $C = \frac{\beta}{g_s l_s} \sqrt{2 \pi L_9^{\ast} \beta}$ \\
Using, $\prod_{n = 1, odd}^{\infty} C = C^{\zeta_{odd} (0)} = 1$, \  $\prod_{n
 = 1, odd}^{\infty} n = e^{-\zeta_{odd}' (0)} = \sqrt{2}$ \ and rearranging the products we get,
\begin{equation}
  Z_{F 1} = \sqrt{2} \prod_{n = 1, odd}^{\infty} \prod_{m = 1}^{\infty} (1 +
  \frac{( \frac{\pi^2 L_9^{\ast} n}{\beta})^2}{\pi^2 m^2})
\end{equation}
Using, $\frac{\sinh (x)}{x} = \prod_{k = 1}^{\infty} (1 + \frac{x^2}{\pi^2
n^2})$ we get,
\begin{eqnarray}
  Z_{F 1} & = & \prod_{n = 1, odd}^{\infty} \sinh ( \frac{\pi^2 L_9^{\ast}
  n}{\beta}) \nonumber\\
  & = & \frac{\prod_{n = 1}^{\infty} 2 \sinh ( \frac{\pi^2 L_9^{\ast}
  n}{\beta})}{\prod_{n = 1}^{\infty} 2 \sinh ( \frac{2 \pi^2 L_9^{\ast}
  n}{\beta})} \nonumber\\
  & = & \frac{\eta (i \frac{\pi L_9^{\ast}}{\beta})}{\eta (i \frac{2 \pi
  L_9^{\ast}}{\beta})} \nonumber\\
  & = & \frac{\eta (i \frac{1}{2 x})}{\eta (i \frac{1}{x})} 
\end{eqnarray}
Where $x = \frac{\beta}{2 \pi L_9^{\ast}}$, and using property of Dedekind eta
function,we get,
\begin{equation}
  Z_{F 1} = \sqrt{2} \frac{\eta (2 ix)}{\eta (ix)}
\end{equation}
\section{Review of Background Gauge Fixing Method} 
Consider dimensionally reduced Maximally Super-symmetric Yang-Mills theory in 10 dimensions 
to 2 dimension. \ The Lagrangian is given by,
\begin{equation}
  L = \frac{1}{g_{YM}^2} Tr (- (D_{\mu} A^i)^2 + \theta^T D \hspace{-0.08in} /
  \theta - \frac{1}{2} F_{\mu \nu}^2 - \frac{1}{2} \lbrack A^i, A^j \rbrack^2 + \theta^T
  \gamma_i \lbrack A^i, \theta \rbrack )
\end{equation}
Where $i = 1, ...8$ and $\mu, \nu = 0, 9$. \ The metric is $\eta_{\mu, \nu} = (-
1, 1, 1..., 1)$. \  $\theta$ is 16 component Maiorana Spinor and $\gamma$
matrices obey 10-dimensional Clifford Algebra. \ Let us consider,
\begin{eqnarray}
  A_{\mu} & = & a_{\mu} + A_{\mu}' \nonumber\\
  A_i & = & a_i + A_i' \nonumber\\
  \theta & = & \Theta + \theta' 
\end{eqnarray}
Where $a_{\mu}, a_i, \Theta$ are background fields obeying classical equation
of motion. \ Let us define a new covariant derivative as $\bar{D}_{\mu} =
\partial_{\mu} + ia_{\mu}$. \ The primed fields are quantum fluctuations which is
integrated out for calculation of partition function. \ Also,
\begin{equation}
  F_{\mu \nu} = \bar{F}_{\mu \nu} + ( \bar{D}_{\mu} A_{\nu}' -
  \bar{D}_{\nu} A_{\mu}') + i \lbrack A_{\mu}', A_{\nu}' \rbrack
\end{equation}
where, $\bar{F}_{\mu \nu} = \partial_{\mu} a_{\nu} - \partial_{\nu}
a_{\mu} + i \lbrack a_{\mu}, a_{\nu} \rbrack $.\\
Now the allowed gauge transformation is the ones which keep the background
unchanged, i.e. $\delta a_{\mu} = \delta a_i = \delta \Theta = 0$. \ Then the
gauge transformation on the fluctuations are given by,
\begin{eqnarray}
  \delta A_{\mu}' & = & \bar{D}_{\mu} \alpha + i \lbrack A_{\mu}', \alpha \rbrack
  \nonumber\\
  \delta A_i' & = & i \lbrack A_i', \alpha \rbrack \nonumber\\
  \delta \theta' & = & i \lbrack \theta', \alpha \rbrack
\end{eqnarray}
Let us choose $a_i = 0$ and $\Theta = 0$.\\
The gauge fixing condition we use is,
\begin{equation}
  \bar{D}_{\mu} A^{\mu \prime} = 0
\end{equation}
therefore the Gauge fixing Lagrangian,
\begin{equation}
  L_{gf} = - \frac{1}{2 g_{YM}^2} Tr ( \bar{D}_{\mu} A^{\mu \prime})^2
\end{equation}
and the ghost Lagrangian,
\begin{equation}
  L_{gh} = Tr ( \bar{\omega} \bar{D}_{\mu} \bar{D}^{\mu} \omega + i
  \bar{\omega} \bar{D}_{\mu} \lbrack A^{\mu \prime}, \omega \rbrack )
\end{equation}
Where $\omega$ and $\bar{\omega}$ are ghost and anti ghost respectively. \ The
background for ghost fields are taken to be zero.

Now we calculate the Lagrangian up to 1-loop level,i.e. we keep terms up to
quadratic in fluctuations. \ And we also use classical equation of motion for
background fields. \ We get,
\begin{eqnarray}
  L_{1 loop}&=& \frac{1}{2 g_{YM}^2} Tr ( A^{\prime i} \bar{D}^2 A^{\prime i} -
  A_0' \bar{D}^2 A_0' + A_9' \bar{D}^2 A_9' + \theta^{\prime T} \bar{D}
  \hspace{-0.1in} / \theta' \nonumber \\ 
  &-& \frac{1}{2} \bar{F}_{\mu \nu}^2 - i
  \bar{F}^{\mu \nu}  A_{\mu}' A_{\nu}' ) + \bar{\omega} \bar{D}^2 \omega
\end{eqnarray}
Let us choose $a_0 = 0$ and $a_9 = constant$, then $\bar{F}^{\mu \nu} =
0$. \ Also scale $\omega$ properly, we get,
\begin{equation}
  L_{1 loop} = \frac{1}{2 g_{YM}^2} Tr ( A^{\prime i} \bar{D}^2 A^{\prime i} -
  A_0' \bar{D}^2 A_0' + A_9' \bar{D}^2 A_9' + \theta^{\prime T} \bar{D}
  \hspace{-0.1in} / \theta' + \bar{\omega} \bar{D}^2 \omega )
\end{equation}
The Euclidean partition function is given by,
\begin{equation}
  lnZ = \frac{10}{2} Tr (ln \bar{D}^2)_{bosonic} - \frac{16}{4} Tr (ln
  \bar{D}^2)_{fermionic} - Tr (ln \bar{D}^2)_{ghost}
\end{equation}

\end{document}